\documentclass[aps,prb,amssymb,showpacs,twocolumn,longbibliography]{revtex4-2}

\usepackage{graphics,graphicx,amsmath}
\usepackage{tabularx,float}
\usepackage{epsfig}
\usepackage{subfigure}
\usepackage{bm}
\usepackage{stmaryrd}
\usepackage{mathtools}
\usepackage{xcolor}
\usepackage{wasysym}
\usepackage{amsthm}
\usepackage{bbm}
\usepackage[T1]{fontenc}
\usepackage[english,french]{babel}

\usepackage{mathrsfs} 
\usepackage{physics}

\usepackage{color}
\usepackage{verbatim}
\graphicspath{{fig/}}
\newtheorem{proposition}{Proposition}

\DeclareMathOperator{\adj}{adj}

\begin{document}

\date{\today}

\title{$\mathbb{Z}_2$ topological invariants from the Green function's diagonal zeros}

\author{Florian Simon}
\affiliation{Laboratoire de Physique des Solides, Universit\'e Paris Saclay,
CNRS UMR 8502, F-91405 Orsay Cedex, France} 
\author{Corentin Morice}
\affiliation{Laboratoire de Physique des Solides, Universit\'e Paris Saclay,
CNRS UMR 8502, F-91405 Orsay Cedex, France}

\begin{abstract}
       We investigate the relationship between the analytical properties of the Green's function and $\mathbb{Z}_2$ topological insulators, focusing on three-dimensional inversion-symmetric systems. We show that the diagonal zeros of the Green's function in the orbital basis provide a direct and visual way to calculate the strong and weak $\mathbb{Z}_2$ topological invariants. We introduce the surface of crossings of diagonal zeros in the Brillouin zone, and show that it separates time-reversal invariant momenta (TRIMs) of opposite parity in two-band models, enabling the visual computation of the $\mathbb{Z}_2$ invariants by counting the relevant TRIMs on either side. In three-band systems, a similar property holds in every case except when a trivial band is added in the band gap of a non-trivial two-band system, reminiscent of the band topology of fragile topological insulators.
\end{abstract}

\maketitle
\section*{Introduction}

The ubiquitous presence of topological states in matter, and its permeation into much of condensed matter physics \cite{Klitzing1980, Thouless1982, Simon1983, Hasan2010, Fu2011, FKM1, FKM2, FKM3, FKM4, FKM5, Bernevig2006, Hasan2010, Cayssol2013, Sato2017, Ozawa2019, Cayssol2021, Xu2024}, contrasts with the ill-adaptation of its theoretical framework to the physical reality it describes. Indeed, topological band theory remains based on non-interacting Bloch systems, is limited to zero temperature, and cannot fully take into account the effects of electronic correlations. Moreover, direct measurements of topological invariants remain unavailable, forcing the field to rely heavily on ab-initio calculations whose calculated invariants are inaccurate even in some simple cases \cite{Vidal2011, Aguilera2021}.

These limitations motivated early on the calculation of topological invariants in more advanced formalisms \cite{Niu1985}, in particular using Green's functions, which can be closely related to experimental results. Topological field theory yielded expressions which are difficult to use in practice, involving integrals of products of Green's functions \cite{ Volovik1989, Yakovenko1990, Golterman1993, Qi2008, Volovik2009, ZhongWang2010, ZhongWang2010b, Xiao-Liang2011, Gurarie2011, Wang2011, Wang2012, Gurarie2011}, such as the So-Ishikawa-Matsuyama invariant \cite{So1985,Ishikawa1986, Matsuyama1987}. Further work addressed this problem by defining a topological Hamiltonian, equal to the inverse of the full Green's function at zero frequency \cite{ZhongWang2012, Wang2012b, Wang2013, He2016I, He2016II, WitczakKrempa2014}. It can then be used to calculate invariants using the same techniques used for non-interacting systems. Similar approaches involving frequency dependence were also recently developed \cite{Blason2023, Krishtopenko2025}.

Another path to the calculation of topological invariants from Green's functions was opened recently. Research on neutrino oscillations prompted the rediscovery of a mathematical identity relating the eigenvalues of a matrix to its eigenvectors \cite{Denton2020, Denton2021}. In the case of a Green's function matrix, this identity involves the zeros of the diagonal components of the Green's function in the orbital basis. Combining it with a known inequality from Cauchy, it was discovered that, in simple models, these zeros signal the presence of a topological state \cite{Misawa2022}. Such zeros can be accessed by directly reading the Green's function matrix, without any inversion nor diagonalisation, making this approach potentially simpler than the canonical non-interacting techniques when the Green's function can be readily accessed, for example via experimental measurements. Diagonal zeros were also studied in interacting systems in connection with the topological Hamiltonian \cite{Tran2022}.

Different zeros of the Green's function, namely zeros of its determinant, or equivalently of its eigenvalues, are currently garnering a lot of interest. Caused by divergences of the self-energy and indicative of the breakdown of perturbation theory, they have been related to failures of Green's function-based topological invariants \cite{Seki2017, Skolimowski2022, Zhao2023, PeraltaGavensky2023, Bollmann2024}. They can form a surface in reciprocal space called the Luttinger surface \cite{Dzyaloshinskii2003, Skolimowski2022, Zhao2023}, which arises in Mott insulators \cite{Rosch2007,Wagner2023}, and could even elucidate the origin of Fermi arcs in the pseudogap phase of cuprate superconductors \cite{Worm2024}.

In this work, focusing on three-dimensional inversions symmetric systems, we show that zeros of the diagonal components of the Green's function in the orbital basis can be used to calculate the $\mathbb{Z}_2$ topological invariants. In Sec. \ref{Sec:Zéro-topologie}, we briefly review the formalism of diagonal zeros. In Sec. \ref{Sec:two-band-models}, we introduce the concept of zeros surface, formed by points of the Brillouin zone where two zeros of the Green's function cross, and show that in two-band models this surface separates TRIMs of opposite parity, allowing one to visually determine the value of $\mathbb{Z}_2$ topological invariants. In Sec. \ref{Sec:Wilson-Dirac}, we exemplify our findings in a generic toy-model for three-dimensional $\mathbb{Z}_2$ topological insulators. Finally, in Sec. \ref{Sec:Counter-Example}, we treat the case of three-band systems, and find that the zeros surface can be affected by the addition of trivial bands.

\section{Diagonal zeros and topology}
\label{Sec:Zéro-topologie}
For completeness, in this section we review the formalism of the zeros of the diagonal components of the Green's function in the orbital basis which we call diagonal zeros in the following \cite{Denton2021,Misawa2022}, with revised notations. We then discuss the expression of the Green's function diagonal components in spin-degenerate band structures and formulate how the diagonal zeros may give access to the value of the $\mathbb{Z}_2$ topological invariants.

\subsection{Notation}
\label{Ssec:Notations}
 Let $\mathcal{M}_N(\mathbb{C})$ be the set of $N$ by $N$ complex matrices. For a Hermitian matrix $A\in\mathcal{M}_N(\mathbb{C})$, we consider the convention in which its eigenvalues are arranged in ascending order, namely $\lambda_1(A)\leq\lambda_2(A)\leq\cdots\leq\lambda_N(A)$.  $[A]_{ij}\in\mathcal{M}_{N-1}(\mathbb{C})$ denotes the matrix $A\in\mathcal{M}_N(\mathbb{C})$ without its $i$-th row and $j$-th column. The eigenvalues of the Hamiltonian $H$ and $[H]_{ij}$ are denoted as $\epsilon_n$ and $\zeta_{n,ij}$, respectively. Finally, for lightness of notation, we will use $\zeta_{nj}$ instead of $\zeta_{n,jj}$. As we consider systems with both inversion and time-reversal symmetry, the band dispersions are spin-degenerate and we show in Sec. \ref{Ssec:Zero-TRIS} that the spin degree of freedom is not relevant for our discussion.
 We therefore consider a spinless Bloch Hamiltonian $H$ represented in an orbital basis $(\ket{\phi_1},\hdots,\ket{\phi_N})$ \footnote{As in Ref. \cite{Törma2023}, we refer to any collection of internal degrees of freedom (spin, orbital, sublattice,..) as an orbital.}. In this basis, the Bloch state of the $n$-th band $\ket{u_n}$ reads
\begin{equation}
    \ket{u_n(\bm{k})}=\sum_{j=1}^Nu_n^{(j)}(\bm{k})\ket{\phi_j},
    \label{eq:Bloch-state-super}
\end{equation}
with $u_n^{(j)}(\bm{k})=\braket{\phi_j}{u_n(\bm{k})}\in\mathbb{C}$. In this paper we consider the situation of a tight-binding Hamiltonian, where the orbital states refer to the on-site atomic orbitals, along with the possible sublattice degree of freedom. The orbital parities then correspond to the parity of the atomic orbital.
\subsection{Formalism}
\label{Ssec:Formalism}
We review the formalism of the diagonal zeros, focusing on its three main relations: The zero-pole representation of the Green's function, the eigenvector-eigenvalue relation, and the Cauchy interlacing inequality.
\subsubsection{Zero-pole representation of $G$}
\label{Sssec:Zero-pole-rep}
Let $H$ be a quadratic Hamiltonian. Its retarded Green's function in frequency space is given by $G=\big(\hbar\omega\mathbbm{1}-H\big)^{-1}$. Using the relation $A\adj(A)=\det(A)\mathbbm{1}$ with $\adj(A)_{ij}=(-1)^{i+j}\det [A]_{ji}$ the adjugate matrix of $A$, one gets
\begin{equation}
    G_{ij}=(-1)^{i+j}\frac{\det\big([\hbar\omega\mathbbm{1}-H]_{ji}\big)}{\det\big(\hbar\omega\mathbbm{1}-H\big)},
    \label{eq:Gij}
\end{equation}
 Diagonal components then simplify to 
\begin{equation}
    G_{jj}=\frac{\det\big(\hbar\omega\mathbbm{1}-[H]_{jj}\big)}{\det\big(\hbar\omega\mathbbm{1}-H\big)}=\frac{\overset{N-1}{\underset{n=1}{\prod}}\big(\hbar\omega-\zeta_{nj}\big)}{\overset{N}{\underset{n=1}{\prod}}\big(\hbar\omega-\epsilon_n\big)}.
    \label{eq:Gjj-zero-pole-rep}
\end{equation}
We will refer to Eq.~(\ref{eq:Gjj-zero-pole-rep}) as the \textit{zero-pole representation} of $G$. $\zeta_{nj}$ then appears as the $n$-th zero of the $j$-th diagonal component of $G$. As such, we refer to them as the \textit{diagonal zeros} of $G$. We emphasize the difference between the diagonal zeros, those of $G_{jj}$ with $G$ written in the orbital basis, and the \textit{singular zeros} of $G$, those of $\det G$ that appear in correlated systems due to divergences of the self-energy \cite{Volovik2009,Seki2017,Skolimowski2022,Zhao2023,Wagner2023,Blason2023}. Points in the Brillouin zone where a diagonal zero $\zeta_{nj}$ coincides with a pole $\epsilon_m$ will hereafter be referred to as \textit{zero-pole touchings}. At such points, a zero of $G_{jj}$ cancels out a pole, making the associated band invisible to spectroscopic techniques. Such situations appear in quantum chaos and gauge-gravity duality, where they are are referred to as pole-skipping points \cite{Wu2019,Grozdanov2019,Ceplak2020,Natsuume2020,Ahn2021}.

 The diagonal zero $\zeta_{nj}$, as the eigenenergy of $[H]_{jj}$, can be interpreted as a \textit{marginal band dispersion}. Indeed, $[H]_{jj}$ being the Hamiltonian without the degree of freedom $\ket{\phi_j}$, the $\zeta_{nj}$ are what the band dispersion $\epsilon_n$ would be \textit{without} the degree of freedom $\ket{\phi_j}$, analogously to marginal probability distributions. Such distributions are obtained by integrating out, i.e. marginalizing, unwanted variables to focus on others. Thus, we can identify these points to the zeros of $G_{jj}$ with respect to $\hbar\omega$. The function $G_{jj}$, being the Fourier transform of the $\ket{\phi_j}\hspace{-1mm}-\hspace{-1mm}\ket{\phi_j}$ response function, has zeros at the energies where the orbital $\ket{\phi_j}$ has no influence in the band structure. 

\subsubsection{Eigenvector-eigenvalue relation}
\label{Sssec:Eig-val-vec-rel}
Another central relation is the \textit{eigenvector-eigenvalue relation} \cite{Denton2021}, which in band theory notation is given by
\begin{equation}
    \prod_{m=1}^{N-1}\big(\epsilon_n-\zeta_{mj}\big)=\big|u_n^{(j)}\big|^2\prod_{\substack{m=1\\m\neq n}}^N\big(\epsilon_n-\epsilon_m\big).
    \label{eq:vec-val-relation}
\end{equation}
 Eq.~(\ref{eq:vec-val-relation}) implies that the zero-pole touchings coincide with the zeros of the Bloch state components, assuming no band crossing. More specifically, the $j$-th component of the $n$-th Bloch state is zero when one diagonal zero of $G_{jj}$ coincides with the $n$-th pole (or dispersion) $\epsilon_n$. A generalised version of the eigenvector-eigenvalue relation can be used to extract the relative phase between $u_n^{(i)}$ and $u_n^{(j)}$,
\begin{equation}
    \exp\Big[i(\phi_n^{(i)}-\phi_n^{(j)})\Big]=\frac{(-1)^{i+j}\det\big[\epsilon_n\mathbbm{1}-H\big]_{ji}}{\sqrt{\overset{N-1}{\underset{m=1}{\prod}}\big(\epsilon_n-\zeta_{mi}\big)\big(\epsilon_n-\zeta_{mj}\big)}},
    \label{phase relative}
\end{equation}
where we have $u_n^{(j)}=|u_n^{(j)}|\text{exp}\big(i\phi_n^{(j)}\big)$. The presence of differences between zeros and poles in the denominator connects zero-pole touchings to phase singularities in the Bloch states' components, where the phase is ill-defined. It in turn also connects the diagonal zeros to the phase's winding numbers, which add up to the Chern number in two-dimensional systems \cite{Hatsugai1993,Hatsugai1997,Bernevig2013,Funfhaus2022,Gunawardana2024}. For completeness, we provide proofs for Eqs. (\ref{eq:vec-val-relation},\ref{phase relative}) \cite{SM}. 

The fact that in the absence of band degeneracies the zeros of the Bloch states components are equivalent to zero-pole touchings can be further explained using the interpretation of diagonal zeros. Indeed, we interpreted $\zeta_{nj}$ as a marginal band dispersion, when the orbital $\ket{\phi_j}$ is disregarded. It then follows that a point $\bm{k}$ such that $\zeta_{nj}(\bm{k})$ is equal to $\epsilon_n(\bm{k})$ is a point where the orbital $\ket{\phi_j}$ does not contribute to the Bloch state $\ket{u_n(\bm{k})}$. In other words, at such points we have $u_{n}^{(j)}(\bm{k})=0$. Such zeros of Bloch state's components were, before the connection was made with Green's function's zeros, already linked to properties of topological semimetals \cite{Piechon2013}.

\subsubsection{Cauchy interlacing inequality}
\label{Sssec:Cauchy-inter-ineq}
The last fundamental relation is the Cauchy interlacing inequality \cite{Denton2021,Fisk2005}, stating that the zeros of $G_{jj}$ interlace its poles, i.e.
\begin{equation}
    \epsilon_n\leq\zeta_{nj}\leq\epsilon_{n+1},
    \label{eq:Cauchy-inequality}
\end{equation}
restricting the possible zero-pole touchings. Indeed, applying this to the eigenvector-eigenvalue relation yields 
\begin{equation}
    u_n^{(j)}=0\hspace{2mm}\Rightarrow\hspace{2mm}\zeta_{n-1,j}=\epsilon_n\hspace{2mm}\text{or}\hspace{2mm}\zeta_{nj}=\epsilon_n.
    \label{eq:zéros-coeff-zéros-diag}
\end{equation}
For completeness, we provide a proof of the Cauchy interlacing inequality \cite{SM}. 

\subsection{Inversion and time-reversal symmetric systems}
\label{Ssec:Zero-TRIS}
 In this subsection, we consider an  inversion and time-reversal symmetric spinful Hamiltonian $H(\bm{k})$ with $2N$ bands, such that no (asymmetric) spin-orbit coupling is present and spin is a good quantum number. The Hamiltonian is therefore a direct sum of the $N$-band spin-up and spin-down Hamiltonians, respectively denoted as $h^{\uparrow}(\bm{k})$ and $h^{\downarrow}(\bm{k})$. Time-reversal symmetry implies that $h^{\uparrow}(\bm{k})=h^{\downarrow}(-\bm{k})^*$ while inversion symmetry implies $h^{\sigma}(\bm{k})=h^{\sigma}(-\bm{k})$, for $\sigma\in\{\uparrow,\downarrow\}$. Systems presenting both symmetries therefore obey $h^{\uparrow}(\bm{k})=h^{\downarrow}(\bm{k})^*$. This implies that for $1\leq n\leq N$ we have $\epsilon_{n\uparrow}=\epsilon_{n\downarrow}$, i.e. the band dispersions of $H$ are spin-degenerate. Additionally,  let $\zeta_{nj}^{\sigma}$ be the marginal band dispersions of $h^{\sigma}$, i.e. the eigenvalues of the matrix $h^{\sigma}$ without the basis state $\ket{\phi_j\sigma}$, denoted $[h^{\sigma}]_{jj}$. Since $[h^{\sigma*}]_{jj}=[h^{\sigma}]_{jj}^*$, we have $[h^{\uparrow}]_{jj}=[h^{\downarrow}]_{jj}^*$ which also implies that $\zeta^{\uparrow}_{nj}=\zeta^{\downarrow}_{nj}$. The marginal band dispersions are therefore also spin-degenerate. 

 The Green's function $G\in\mathcal{M}_{2N}(\mathbb{C})$ written in the orbital basis $\ket{\phi_j\sigma}$ has diagonal elements given by
 \begin{equation}
     G_{j\sigma,j\sigma}=\frac{\det\big(\hbar\omega\mathbbm{1}-[H]_{j\sigma,j\sigma}\big)}{\det(\hbar\omega\mathbbm{1}-H)},
 \end{equation}
 where $[H]_{j\sigma,j\sigma}$ the spinful Hamiltonian $H$ without the basis state $\ket{\phi_j\sigma}$. Since $H$ is the direct sum of $h^{\sigma}$ and $h^{\overline{\sigma}}$, with $\overline{\sigma}$ the opposite spin value of $\sigma$, the submatrix $[H]_{j\sigma,j\sigma}$ is given by $[H]_{j\sigma,j\sigma}=[h^{\sigma}]_{jj}\oplus h^{\overline{\sigma}}$. The diagonal element of the Green's function then becomes
 \begin{subequations}
     \begin{align}
         G_{j\sigma,j\sigma}&=\frac{\overset{N-1}{\underset{n=1}{\prod}}(\hbar\omega-\zeta^{\sigma}_{nj})\overset{N}{\underset{n=1}{\prod}}(\hbar\omega-\epsilon_{n\overline{\sigma}})}{\overset{N}{\underset{n=1}{\prod}}(\hbar\omega-\epsilon_{n\sigma})(\hbar\omega-\epsilon_{n\overline{\sigma}})}\\
         &=\frac{\overset{N-1}{\underset{n=1}{\prod}}(\hbar\omega-\zeta^{\sigma}_{nj})}{\overset{N}{\underset{n=1}{\prod}}(\hbar\omega-\epsilon_{n\sigma})}.
     \end{align}
 \end{subequations}
 The spin-degeneracy of the band dispersions and the marginal band dispersions then directly shows that $G_{j\uparrow,j\uparrow}=G_{j\downarrow,j\downarrow}\coloneqq G_{jj}$. An explicit example is seen in Sec. \ref{Sec:Wilson-Dirac}, with the Wilson-Dirac model \cite{Yoshimura2016}.

\subsection{$\mathbb{Z}_2$ topological invariants}
\label{Ssec:General-prop}
By exploring a variety of toy models in several classes of the tenfold way \cite{Schnyder2008,Kitaev2009,Ryu2010,Chiu2016}, the authors of Ref. \cite{Misawa2022} discovered that upon a band inversion, zero-pole touchings can undergo rearrangements, such that two diagonal zeros $\zeta_{ni}$ and $\zeta_{nj}$ cross in the gap between two inverted zero-pole touchings. The presence of these \textit{zero-zero crossings} in the Brillouin zone then signals band inversions. In the case of non-trivial band inversions, where the involved orbitals have different orbital parities, the zero-zero crossings signal the topological non-triviality of the band structure.
One can wonder if it is possible to go further and extract the value of topological invariants from the diagonal zeros of the Green's function. In the following, we formulate a proposition to do so in $\mathbb{Z}_2$ topological systems. Specifically, we focus on time-reversal and inversion symmetric three-dimensional topological insulators. Our reasoning can readily be adapted to systems in other dimensions. In the three dimensional case, the four $\mathbb{Z}_2$ topological invariants $(\nu_0;\nu_1\nu_2\nu_3)$ are given by the Fu-Kane-Mele formula \cite{FKM4,FKM5}
\begin{equation}
    (-1)^{\nu_j}=\prod_{\Gamma_i\in\text{T}_j}\xi(\Gamma_i)=\prod_{\Gamma_i\in\text{T}_j}\prod_{n=1}^{N_{\text{occ}}}\xi_n(\Gamma_i),
    \label{eq:FKM-Z2}
\end{equation}
where $\xi(\Gamma_i)$ and $\xi_n(\Gamma_i)$ are the band parity of the TRIM $\Gamma_i$ and that of the $n$-th band at $\Gamma_i$, respectively. Additionally, $\text{T}_j$ is a subset of TRIMs in the Brillouin zone (BZ). More specifically, $\text{T}_0$ contains all eight TRIM and the other three sets $\text{T}_j=\big\{n_{j}=1;n_{i\neq j}=0,1\big\}$ contain four TRIMs, where the $n_j$ are the indices of the TRIMs, such that $\Gamma_i=\sum_jn_j/2\bm{b}_j$. The latter are pictured in Fig. \ref{fig:Figure-Deux-Bandes}a in the case of the cubic Brillouin zone. Finally, $N_{\text{occ}}$ is the number of occupied bands.

It is then readily shown from Eq.~(\ref{eq:FKM-Z2}) that
\begin{equation}
    \nu_j=|\mathcal{P}_{j,\pm}|\mod2,
    \label{eq:Invariant-Parité-Secteur}
\end{equation}
with $\mathcal{P}_{j,\pm}\subset\text{T}_j$ the set of TRIM in $\text{T}_j$  with parity $\pm1$, and $|\mathcal{P}_{j,\pm}|$ its cardinality. The topological invariant $\nu_j$ is therefore equal to the number of TRIM with the same parity, modulo 2. The invariants $\nu_j$ can then be obtained by counting the number of TRIM in $\text{T}_j$ within the same parity sector. 
Since they signal band inversions, diagonal zeros can be used to delimit the parity sectors. 
\section{Two-band models}
\label{Sec:two-band-models}
We now prove that the $\mathbb{Z}_2$ invariants are determined by the diagonal zeros in the case of two-band models. We consider a Hamiltonian $H$ written in the orbital basis $(\ket{\phi_1},\ket{\phi_2})$ with orbital parities $(\xi_1,\xi_2)$ respectively. The two Bloch states $\ket{u_1}$ and $\ket{u_2}$ have, at a TRIM $\Gamma_i$, band parities $\xi_1(\Gamma_i)$ and $\xi_2(\Gamma_i)$, respectively. Furthermore we consider half-filling, i.e $N_{\text{occ}}=1$. We start by formally articulating the central claim, being that the surface in the Brillouin zone formed by the zero-zero crossing separates TRIMs of opposite parity, which we then prove by establishing the equivalence between band inversions and zero-zero crossings. We then underscore the importance of the non-triviality of this surface, referred to as the zeros surface in the following.
\subsection{Proposition}
\label{Ssec:two-band-general-form}
 The key point is to prove that TRIMs with opposite parity are separated by zero-zero crossings. Pursuing this line of thought, let $\Gamma_1$ and $\Gamma_2$ be two TRIM points in $\text{T}_j$. Their parities are then given by $\xi(\Gamma_{1,2})=\xi_1(\Gamma_{1,2})$. For the diagonal zeros $\zeta_{11}$ and $\zeta_{12}$, we define the zeros surface as follows,
\begin{equation}
    \mathscr{S}_0=\Big\{\bm{k}\in\text{BZ}\hspace{1mm}\Big|\hspace{1mm}\zeta_{11}(\bm{k})=\zeta_{12}(\bm{k})\Big\},
    \label{eq:Two-bands-Surface-zéros}
\end{equation}
i.e. as the set of points in the Brillouin zone where the two diagonal zeroes cross. We will refer to the zeros surface as being non-trivial iff the two orbitals $\ket{\phi_1}$ and $\ket{\phi_2}$ are of opposite orbital parities, i.e. $\xi_1\neq\xi_2$, in direct analogy with the non-triviality of band inversions. Otherwise, it is called trivial. Our proposition is then written as
\begin{proposition}
    Two TRIMs $\Gamma_1$ and $\Gamma_2$ are of opposite parities $\xi(\Gamma_1)$ and $\xi(\Gamma_2)$ iff the non-trivial zeros surface $\mathscr{S}_0$ separates $\Gamma_1$ and $\Gamma_2$
    \label{Prop:two-band-prop}
\end{proposition}

We emphasize the importance of the \textit{non-triviality of the zeros surface}. If the band inversion between $\ket{\phi_1}$ and $\ket{\phi_2}$ is trivial, then $\Gamma_1$ and $\Gamma_2$ will always be of equal parity, and the zeros surface will separate the two TRIMs regardless. This point is further discussed in Sec. \ref{Ssec:Two-bands-parities}. A crossing of trivial zeros could be related to surface states, whose origin would not be topological \cite{Silva2021}. In Section \ref{Ssec:Formalism} we have interpreted the diagonal zeros as marginal band dispersions, corresponding to the disregard of one orbital. Here $\zeta_{11}$, the lowest zero of $G_{11}$, is associated to the omission of $\ket{\phi_1}$ while $\zeta_{12}$ comes from that of $\ket{\phi_2}$. The point $\bm{k}\in\text{BZ}$ where $\zeta_{11}(\bm{k})=\zeta_{12}(\bm{k})$ is thus a point where removing one orbital is equivalent to removing another. 

\subsection{Proof}
\label{Ssec:two-band-proof}
\begin{figure*}[t]
    \centering
    \includegraphics[width=\textwidth]{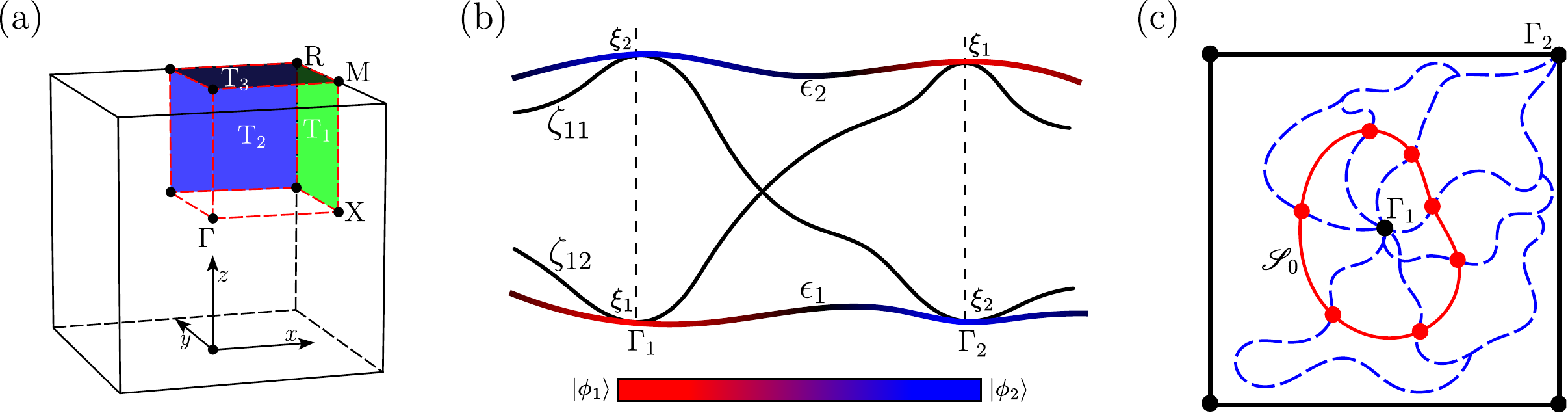}
    \caption{ (a) Visualization of the subsets of TRIMs $\text{T}_j$ corresponding to the three $\mathbb{Z}_2$ topological invariants $\nu_j$, in the cubic Brillouin zone. The set $\text{T}_0$, corresponding to the strong invariant $\nu_0$, comprises the eight TRIM points. (b) Graphical representation of a band inversion between two TRIM in a two-band system. The inset represents the orbital character of the bands, red being purely $\ket{\phi_1}$ and blue purely $\ket{\phi_2}$. (c) Any path (dashed blue lines) between non-trivially inverted TRIMs $\Gamma_1$ and $\Gamma_2$ crosses the zeros surface $\mathscr{S}_0$ (in red), thus separating the two TRIMs. For illustration, a two-dimensional cut of the cubic Brillouin zone is shown.}
    \label{fig:Figure-Deux-Bandes}
\end{figure*}

We now prove Proposition \ref{Prop:two-band-prop} for two-band models. We do so through the notion of \textit{band inversion}. We remind the reader that a band inversion, in two-band models, is characterized by a reversal of characters of the two bands between two points in the Brillouin zone, here $\Gamma_1$ and $\Gamma_2$. Let us consider the general inverted band structure shown in Fig. \ref{fig:Figure-Deux-Bandes}b. The two Bloch states $\ket{u_1(\bm{k})}$ and $\ket{u_2(\bm{k}})$ are, by Eq.~(\ref{eq:Bloch-state-super}), a $\bm{k}$-dependent superposition of the orbital states $\ket{\phi_1}$ and $\ket{\phi_2}$. On the one hand, the lower band will purely be of character $\ket{\phi_1}$ at $\Gamma_1$, meaning $u_1^{(2)}(\Gamma_1)=0$. This first implies, through the eigenvector-eigenvalue relation stated in Eq.~(\ref{eq:vec-val-relation}), that $\zeta_{12}(\Gamma_1)=\epsilon_1(\Gamma_1)$, and second that $\xi_1(\Gamma_1)=\xi_1$ \footnote{This can be formally proven by utilizing $\xi_n(\Gamma_i)=\bra{u_n(\Gamma_i)}P\ket{u_n(\Gamma_i)}$ with $P$ the parity operator, and the closure relation for the orbital basis. The two then yield $\xi_n(\Gamma_i)=\sum_{j=1}^{N}\xi_j|u_n^{(j)}(\Gamma_i)|^2$}. On the other hand, at $\Gamma_2$ the lower band will be purely of character $\ket{\phi_2}$, meaning $u_1^{(1)}(\Gamma_2)=0$. This first implies that $\zeta_{11}(\Gamma_2)=\epsilon_1(\Gamma_2)$, and second that $\xi_1(\Gamma_2)=\xi_2$. The situation is reversed for the higher band. At $\Gamma_1$ it is solely of character $\ket{\phi_2}$, so that $\zeta_{11}(\Gamma_1)=\epsilon_2(\Gamma_1)$ and $\xi_2(\Gamma_1)=\xi_2$. And at $\Gamma_2$ it is purely of character $\ket{\phi_1}$ so that $\zeta_{12}(\Gamma_2)=\epsilon_2(\Gamma_2)$ and $\xi_2(\Gamma_2)=\xi_1$.

Based on this situation, we now prove the equivalence between zero-zero crossings and band inversions. First, as described above and in Fig. \ref{fig:Figure-Deux-Bandes}b, the band inversion results in the two diagonal zeros fully crossing the gap in opposite directions, following what the band is not, in terms of orbital character. Being continuous functions, they must therefore cross at some point between $\Gamma_1$ and $\Gamma_2$. The band inversion therefore implies a zero-zero crossing, i.e. the separation of two inverted TRIMs by the zeros surface $\mathscr{S}_0$. 

Conversely, assume that there is a zero-zero crossing between $\Gamma_1$ and $\Gamma_2$. We distinguish two possibilities, depending on whether the zero-zero crossing follows from a full crossing of the gap by the diagonal zeros or not. If the zero-zero crossing results from a full crossing of the gap, where each diagonal zero coincides with one band dispersion at $\Gamma_1$ and with the other at $\Gamma_2$, then the eigenvector-eigenvalue relation directly implies a band inversion. Indeed, applying Eq.~(\ref{eq:vec-val-relation}) to the four zero-pole touchings yields the associated zeros of the Bloch states' components and thereby shows the band inversion between $\ket{\phi_1}$ and $\ket{\phi_2}$. The possibility of so-called \textit{accidental} zero-zero crossings, that do not result from a full crossing of the gap, was discussed in Ref. \cite{Misawa2022}. Although they can be present, it was also argued that these crossings can be removed by deforming the Hamiltonian without closing the band gap, i.e. staying in the same topological phase \cite{Misawa2022}. Even without removing them, these accidental crossings are distinguishable from the ones associated with band inversions by probing the position of the diagonal zeros at $\Gamma_1$ and $\Gamma_2$. These accidental zero crossings are therefore a possibility which we neglect in the following. Without loss of generality, the diagonal zeros therefore behave as in Fig. \ref{fig:Figure-Deux-Bandes}b. The second possibility being discarded, a zero-zero crossing between $\Gamma_1$ and $\Gamma_2$ therefore implies a band inversion between the two points. 

We have thus proven the equivalence between the separation of $\Gamma_1$ and $\Gamma_2$ by the zeros surface $\mathscr{S}_0$, and the band inversion between these two points. The consequence is that any path from $\Gamma_1$ to $\Gamma_2$ will necessarily cross the zeros surface, as pictured in Fig. \ref{fig:Figure-Deux-Bandes}b.

\subsection{Non-triviality} 
\label{Ssec:Two-bands-parities}
Having established the link between band inversions and the zeros surface, we proceed by relating band inversions to the relative parity of the inverted TRIMs. Since $N_{\text{occ}}=1$, the parities of the TRIM points $\Gamma_1$ and $\Gamma_2$ are given by $\xi(\Gamma_1)=\xi_1(\Gamma_1)$ and $\xi(\Gamma_2)=\xi_1(\Gamma_2)$, respectively. From the situation we consider in Fig. \ref{fig:Figure-Deux-Bandes}b, we then see that the relative parity of $\Gamma_1$ and $\Gamma_2$ is given by 
\begin{equation}
    \xi(\Gamma_1)\xi(\Gamma_2)=\xi_1(\Gamma_1)\xi_1(\Gamma_2)=\xi_1\xi_2.
\end{equation}
The inverted TRIMs $\Gamma_1$ and $\Gamma_2$ will then be of opposite parity iff the band inversion happens between two orbitals of opposite parity. This is the non-trivial character of the band inversion. Coming back to the zeros surface, we have shown that it separates the TRIMs that are inverted from one another. If the associated band inversion is non-trivial, then the zeros surface will indeed delimit the two different parity sectors.

\section{Example: Wilson-Dirac model}
\label{Sec:Wilson-Dirac}
In this section we exemplify the results proven in Section \ref{Sec:two-band-models} using the Wilson-Dirac model, a generalisation of the low-energy model of $\text{Bi}_2\text{Se}_3$ and a standard model for $\mathbb{Z}_2$ topological insulators \cite{Yoshimura2016}. We start by briefly setting the model and studying its band structure. We then three different topological phases, and compute the topological invariants from the zeros surface according to Proposition \ref{Prop:two-band-prop}.
    \subsection{Model}
    \label{Ssec:Wilson-Dirac-Model}
        \subsubsection{Hamiltonian and Green's function}
        \label{Sssec:Wilson-Dirac-Hamiltonian}
        The Hamiltonian is given by \cite{Yoshimura2016}
        \begin{equation}
            H=(m-b_{\mu}\cos k_{\mu})\beta+t_{\mu}\sin k_{\mu}\alpha_{\mu},
        \end{equation}
        with $\beta=\tau_z\otimes\sigma_0$ and $\alpha_{\mu}=\tau_x\otimes\sigma_{\mu}$ and where we make use of the Einstein summation convention over $\mu\in\{x,y,z\}$. The Pauli matrices $\tau_{\mu}$ and $\sigma_{\mu}$ act on the orbital and spin degrees of freedom, respectively.
        The four band dispersions are double degenerate:
        \begin{equation}
            \epsilon_{1,2}=-\epsilon_{3,4}=-\sqrt{(m-b_{\mu}\cos k_{\mu})^2+t_\mu^2\sin^2 k_{\mu}}.
        \end{equation}
        As for the diagonal zeros, since the model has four bands with the spin degree of freedom, we have $12$ diagonal zeros. However, since $\epsilon_1=\epsilon_2$ and $\epsilon_3=\epsilon_4$, the Cauchy interlacing inequality in Eq.~(\ref{eq:Cauchy-inequality}) implies that $\zeta_{1j}=\epsilon_2$ and $\zeta_{3j}=\epsilon_4$ for $j\in\{1,2,3,4\}$. Finally, for the $\zeta_{2j}$, we find
        \begin{equation}
            \zeta_{21}=\zeta_{22}=b_{\mu}\cos k_{\mu}-m=-\zeta_{23}=-\zeta_{24}.
            \label{eq:Diagonal-Zeros-WDmodel}
        \end{equation}
        Since we have $\zeta_{1j}=\epsilon_1$ and $\zeta_{3j}=\epsilon_3$, the zero-pole representation of the Green's function reads
        \begin{subequations}
        \begin{align}
            G_{jj}&=\frac{(\hbar\omega-\zeta_{1j})(\hbar\omega-\zeta_{2j})(\hbar\omega-\zeta_{3j})}{(\hbar\omega-\epsilon_{1})(\hbar\omega-\epsilon_{2})(\hbar\omega-\epsilon_{3})(\hbar\omega-\epsilon_{4})}\\
            &=\frac{\hbar\omega-\zeta_{2j}}{(\hbar\omega-\epsilon_2)(\hbar\omega-\epsilon_4)}.
            \label{eq:Wilson-Dirac-GF}
        \end{align}
        \end{subequations}
        This is a particular example of the relation shown in Sec. \ref{Ssec:Zero-TRIS}.
        Thus, the only relevant diagonal zeros and band dispersions are the $\zeta_{2j}$ and the $\epsilon_{2,4}$. This reduces our problem to a two-band spinless one.
        \subsubsection{Zeros surface}
        \label{Sssec:Wilson-Dirac-S0}
        As stated in Eq.~(\ref{eq:Two-bands-Surface-zéros}), the zeros surface $\mathscr{S}_0$ is defined by the crossing of two diagonal zeros. Eq.~(\ref{eq:Wilson-Dirac-GF}) shows that only the four diagonal zeros $\zeta_{2j}$ appear non-trivially in $G_{jj}$. Only two of them are not identically degenerate, we will therefore focus on them. We then consider the zero-zero crossings happening between $\zeta_{22}$ and $\zeta_{24}$. Using Eq.~(\ref{eq:Diagonal-Zeros-WDmodel}) then reveals that $\zeta_{22}$ and $\zeta_{24}$ cross for $\bm{k}$ points such that
        \begin{equation}
            b_{x}\cos k_{x}+b_{y}\cos k_{y}+b_{z}\cos k_{z}-m=0.
            \label{eq:Wilson-Dirac-Cond0-1}
        \end{equation}
        As in Ref. \cite{Yoshimura2016}, we now set $b_{x}=b_{y}=b_{\parallel}$ which we assume to be non-zero. Eq.~(\ref{eq:Wilson-Dirac-Cond0-1}) can then be rewritten as
        \begin{equation}
            \cos k_{x}+\cos k_{y}+\frac{b_{z}}{b_{\parallel}}\cos k_{z}=\frac{m}{b_{\parallel}}.
            \label{eq:Wilson-Dirac-Cond0-2}
        \end{equation}
        The zeros surface $\mathscr{S}_0$ is thereby the set of $\bm{k}\in\text{BZ}$ obeying Eq.~(\ref{eq:Wilson-Dirac-Cond0-2}).
        
        \subsection{$\mathbb{Z}_2$ invariants from $\mathscr{S}_0$}
    \label{Ssec:Wilson-Dirac-S0-Phases}
    The topological phase diagram, as a function of $b_{z}/b_{\parallel}$ and $m/b_{\parallel}$, was established in Ref. \cite{Yoshimura2016}. Let us now rederive the $\mathbb{Z}_2$ topological invariants using the zeros surface. Specifically, we will do so in three examples. First, the trivial $(0;000)$ phase, then the weak topological insulator (WTI) $(0;111)$ phase and finally the strong topological insulator (STI) $(1;000)$ phase. The same reasoning extends to any parameter regime of the Wilson-Dirac model.
    \begin{figure*}[t]
        \centering
        \includegraphics[width=\textwidth]{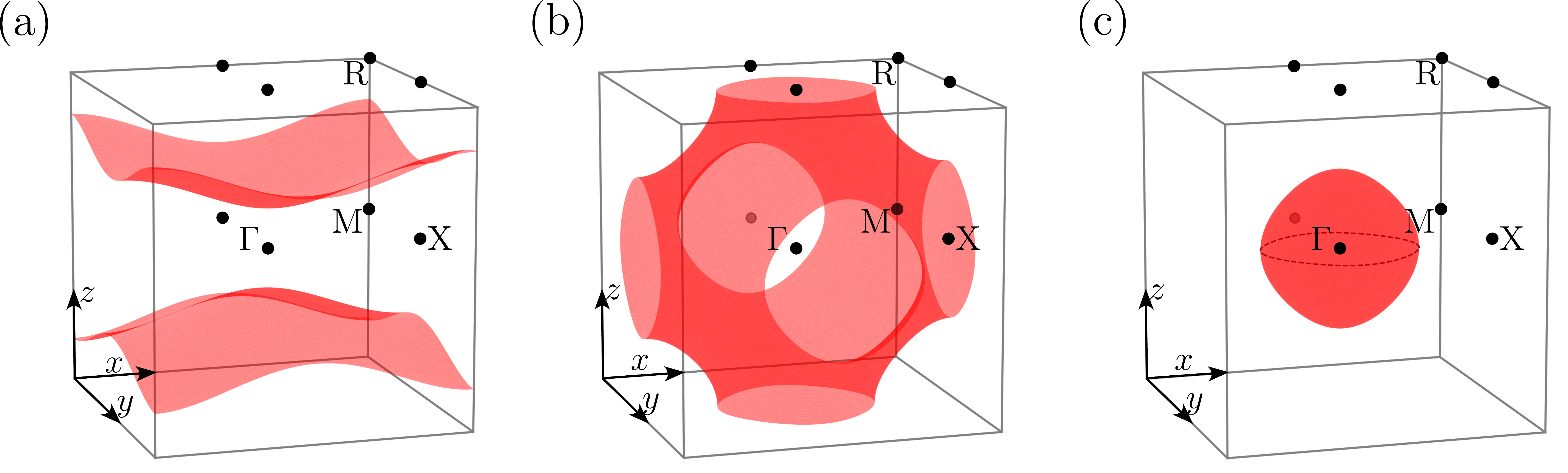}
        \vspace{-5mm}
        \caption{Surface of zero-zero crossings within the cubic Brillouin Zone, for three different topological phases of the Wilson-Dirac model. (a) Zeros surface for the trivial $(0;000)$ phase, with $b_z/b_{\parallel}=-3$ and $m/b_{\parallel}=0$. (b) Zeros surface for the weak $(0;111)$ phase, with $b_z/b_{\parallel}=1$ and $m/b_{\parallel}=0$. (c) Zeros surface for the strong $(1;000)$ phase, with $b_z/b_{\parallel}=1$ and $m/b_{\parallel}=2$.}
        \label{fig:Zeros-Surface-Wilson-Dirac}
    \end{figure*}

        \subsubsection{Trivial phase $(0;000)$}         
        \label{Sssec:Wilson-Dirac-0000}
        For the trivial $(0;000)$ phase, we consider $b_z/b_{\parallel}=-3$ and $m/b_{\parallel}=0$. The resulting zeros surface is shown in Fig. \ref{fig:Zeros-Surface-Wilson-Dirac}a. Using Proposition \ref{Prop:two-band-prop} and Eq.~(\ref{eq:Invariant-Parité-Secteur}), we now recompute the four $\mathbb{Z}_2$ invariants. First, for the strong invariant $\nu_0$, all eight TRIM points of the BZ shown in Fig. \ref{fig:Figure-Deux-Bandes}a are considered. As we can see in Fig. \ref{fig:Zeros-Surface-Wilson-Dirac}a, $\mathscr{S}_0$ separates the four upper TRIM points with the four lower ones. From Eq.~(\ref{eq:Invariant-Parité-Secteur}), we then have $\nu_0= 4\mod2= 0\mod2$, i.e. $\nu_0$ is trivial. We turn to the weak invariants $(\nu_1\nu_2\nu_3)$. First, in computing $\nu_1$, we consider the TRIM points in $\text{T}_1$, shown in Fig. \ref{fig:Figure-Deux-Bandes}a. We then see in Fig. \ref{fig:Zeros-Surface-Wilson-Dirac}a that the four TRIM points are split in two groups of two. Hence each parity sector has two TRIM points, which by Eq.~(\ref{eq:Invariant-Parité-Secteur}) means that $\nu_1$ is trivial. Then, for $\nu_2$, the situation similar to that of $\nu_1$, for the set $\text{T}_2$. So $\nu_2$ is also trivial. Lastly for the weak invariant $\nu_3$, we look at $\text{T}_{3}$. The zeros surface is not present in this plane, therefore each parity sector has an even number of TRIM points, either four or zero. By Eq.~(\ref{eq:Invariant-Parité-Secteur}) we then obtain that $\nu_3$ is trivial. We have thus found, only using the zeros surface, that the phase pictured in Fig. \ref{fig:Zeros-Surface-Wilson-Dirac}a is a trivial $(0;000)$ phase.

        \subsubsection{Weak topological phase $(0;111)$}
        \label{Sssec:Wilson-Dirac-0111}
We now focus on the WTI $(0;111)$ phase, with parameters $b_{z}/b_{\parallel}=1$ and $m/b_{\parallel}=0$. The associated zeros surface is shown in Fig. \ref{fig:Zeros-Surface-Wilson-Dirac}b. For the strong invariant $\nu_0$, we see that we have three X points as well as the $\Gamma_i$ point enclosed by $\mathscr{S}_0$. So an even number of TRIMs are in each parity sector, which implies that $\nu_0$ is trivial. As for the weak invariants $(\nu_1\nu_2\nu_3)$, we see in Fig. \ref{fig:Zeros-Surface-Wilson-Dirac}b that in each case, the X point is separated from the other three TRIM points. By Proposition (\ref{Prop:two-band-prop} and Eq.~(\ref{eq:Invariant-Parité-Secteur}), the three weak invariants $(\nu_1,\nu_2,\nu_3)$ are then non-trivial. From the zeros surface plotted in Fig. \ref{fig:Zeros-Surface-Wilson-Dirac}b, we then correctly infer that the phase is a WTI $(0;111)$.
\subsubsection{Strong topological phase $(1;000)$}
\label{Sssec:Wilson-Dirac-1000}
Finally, we consider the strong $(1;000)$ topological phase with parameters $b_{z}/b_{\parallel}=1$ and $m/b_{\parallel}=2$. The corresponding zeros surface is shown in Fig. \ref{fig:Zeros-Surface-Wilson-Dirac}c. For the strong invariant $\nu_0$ we see in Fig. \ref{fig:Zeros-Surface-Wilson-Dirac}c that the zeros surface encloses the $\Gamma_i$ point, thereby separating it from the other TRIMs. This implies that $\nu_0$ is non-trivial. As for the weak invariants $(\nu_1\nu_2\nu_3)$, we see that in each case the zeros surface does not separate any TRIM from another. By Proposition \ref{Prop:two-band-prop} and Eq.~(\ref{eq:Invariant-Parité-Secteur}), we then conclude that all three weak invariants are trivial. We have thus correctly found from the zeros surface that Fig. \ref{fig:Zeros-Surface-Wilson-Dirac}c is associated to a STI $(1;000)$ phase.

\section{Fragile character}
\label{Sec:Counter-Example}
In Section \ref{Sec:two-band-models} we have established that, within two-band models, the zeros surface separates the inverted TRIMs, thereby allowing for the direct visualization of the $\mathbb{Z}_2$ topological invariants. In this section we explore its possible extension to systems with more bands, hereafter referred to as multiband systems. We begin by formulating a generalisation of Proposition \ref{Prop:two-band-prop} to systems with an arbitrary number of bands. Treating the three-band case in detail then shows that the generalisation holds in all possible cases except when a trivial band is added between two inverted bands, which is reminiscent of fragile topological insulators \cite{Po2018}. 
\subsection{Generalisation of Proposition \ref{Prop:two-band-prop}}
\label{Ssec:N-Bands-General-Prop}
Consider a $N$ band Bloch Hamiltonian represented in the orbital basis $\ket{\phi_1},\hdots,\ket{\phi_N}$.

     The broadest generalisation of Eq.~(\ref{eq:Two-bands-Surface-zéros}) is
    \begin{equation}
        \mathscr{S}_{0,ij}^{n}=\Big\{\bm{k}\in\text{BZ}\hspace{1mm}\Big|\hspace{1mm}\zeta_{ni}(\bm{k})=\zeta_{nj}(\bm{k})\Big\},
        \label{eq:N-bands-S0}
    \end{equation}
    i.e. in every band gap between $\epsilon_n$ and $\epsilon_{n+1}$, each crossing $\zeta_{ni}=\zeta_{nj}$ has its associated zeros surface $\mathscr{S}_{0,ij}^{n}$. In that sense, a $N$-band system will have $N(N-1)$ diagonal zeros and potentially $N(N-1)^2/2$ zeros surface. In this regard, multiband systems are much more complex than their two-band counterparts. We can however motivate that a specific subset of zeros surfaces are relevant to discriminating TRIM points with opposite parities. We expect that only band inversions occurring within the band gap at the Fermi level influence the system's band topology. We can therefore focus only on the zeros surfaces for which $n=N_{\text{occ}}$.  Additionally, we can focus only on pairs $(i,j)$ such that $j>i$, so as to avoid double counting. Finally, it is apparent that $\Gamma_1$ and $\Gamma_2$ are of opposite parity iff an odd number of non-trivial band inversions happen at these points. As in the two-band case, we call the zeros surface $\mathscr{S}_{0,ij}^n$ \textit{non-trivial} iff $\ket{\phi_i}$ and $\ket{\phi_j}$ are of opposite parity. Otherwise, we  call it trivial. Motivated by this, we then propose the following generalisation of Proposition \ref{Prop:two-band-prop}.
\begin{proposition}
    Two TRIM points $\Gamma_1$ and $\Gamma_2$ are of opposite parities $\xi(\Gamma_1)$ and $\xi(\Gamma_2)$ iff they are separated by an odd number of non-trivial zeros surfaces $\mathscr{S}^{N_{\text{occ}}}_{0,ij}$, with $j>i$.
    \label{Proposition-N-bandes}
\end{proposition}
In the following, we find that Proposition 2 is valid in every time-reversal and inversion-symmetric three-band systems except in systems where a (partially) trivial band is present between two inverted bands.
\subsection{Three-band case: fragile zeros surface}
\label{Ssec:Counterexample}
\begin{figure}[t]
    \centering
    \includegraphics[width=\linewidth]{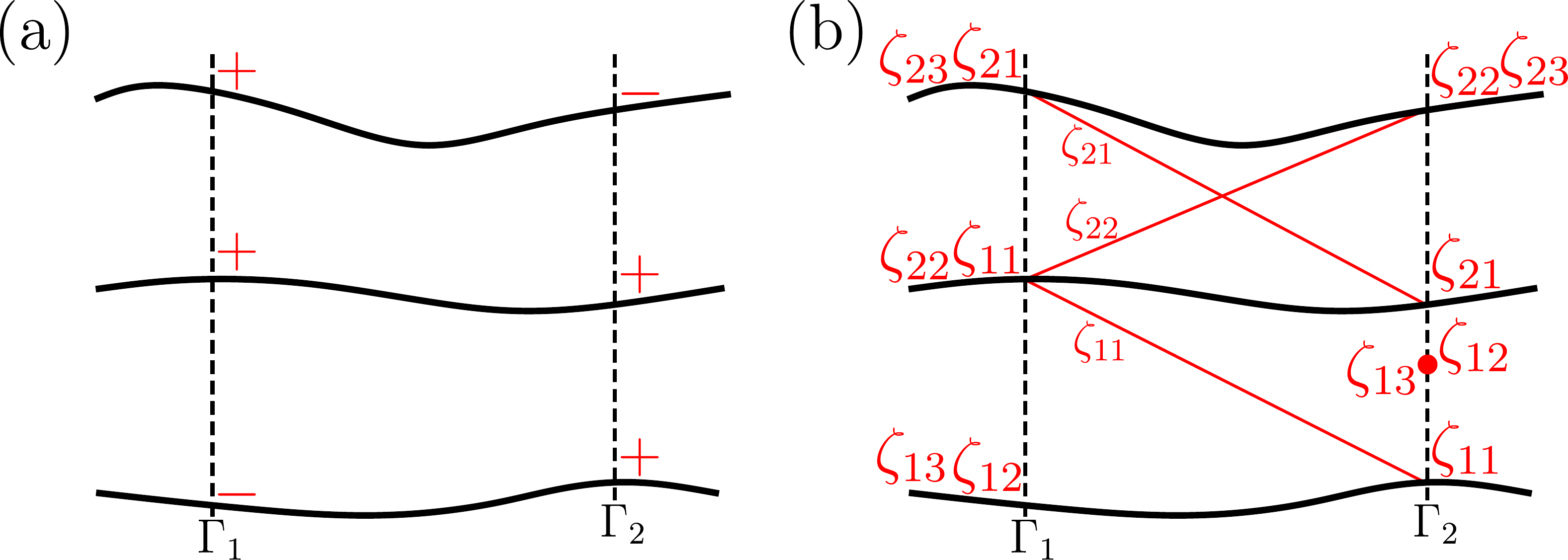}
    
    \caption{ (a) Configuration of band parities for which the Proposition \ref{Proposition-N-bandes} does not hold. (b) Schematic behavior of the diagonal zeros with the configuration of Fig. \ref{fig:Zéros-3band-Gamma1}a at $\Gamma_1$ and of Fig. \ref{fig:Zéros-3band-Gamma2}c at $\Gamma_2$.}
    
    \label{fig:Figure-Trois-Bandes}
\end{figure}

In order to investigate the validity of our results beyond two-models, we computed the positions of the diagonal zeros in all possible three-band cases \cite{SM}. We find that, up to a global sign change of the band parities, Proposition \ref{Proposition-N-bandes} holds in all cases except in one configuration which we display in Fig. \ref{fig:Figure-Trois-Bandes}a. In that configuration, the two TRIMs $\Gamma_1$ and $\Gamma_2$ are of different parity both when $N_{\text{occ}}=1$ and when $N_{\text{occ}}=2$. Proposition \ref{Proposition-N-bandes} then holds if we have a single non-trivial zero crossing between the first and second bands as well as between the second and third ones. We show that the situation of Fig. \ref{fig:Figure-Trois-Bandes}a gives rise to nine different configurations of diagonal zeros, five of which do not obey Proposition \ref{Proposition-N-bandes} \cite{SM}. We now briefly go through their derivations.

\begin{figure}[t]
    \centering
    \includegraphics[width=\linewidth]{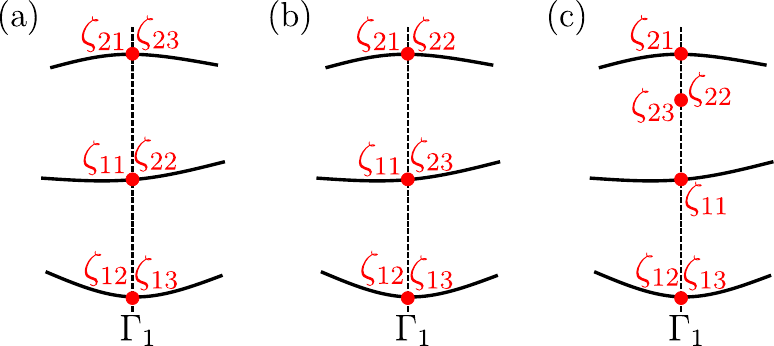
    }
    \caption{The three possible configurations of diagonal zeros at $\Gamma_1$. Figs. \ref{fig:Zéros-3band-Gamma1}a, \ref{fig:Zéros-3band-Gamma1}b and \ref{fig:Zéros-3band-Gamma1}c correspond to the first, second and third solutions in Eq.~(\ref{eq:3band-3sol-Gamma1}), respectively.}
    \label{fig:Zéros-3band-Gamma1}
\end{figure}

 Without loss of generality, we choose the orbital parities to be $(\xi_1,\xi_2,\xi_3)=(-1,+1,+1)$. For a TRIM $\Gamma_i$ let us define the matrix $U(\Gamma_i)$ as follows,
\begin{equation}
    U(\Gamma_i)=\begin{pmatrix}
        u_1^{(1)}(\Gamma_i)&u_2^{(1)}(\Gamma_i)&u_3^{(1)}(\Gamma_i)
        \\
        u_1^{(2)}(\Gamma_i)&u_2^{(2)}(\Gamma_i)&u_3^{(2)}(\Gamma_i)
        \\
        u_1^{(3)}(\Gamma_i)&u_2^{(3)}(\Gamma_i)&u_3^{(3)}(\Gamma_i)
    \end{pmatrix}.
\end{equation}
Using the discrepancy between the orbital and band parities as well as the fact that the orbital and band bases are orthonormal, we have at $\Gamma_1$ 
\begin{equation}
    U=\begin{pmatrix}
        e^{i\phi_1^{(1)}}&0&0\\
        0&\alpha e^{i\phi_2^{(2)}}&\beta e^{i\phi_3^{(2)}}\\
        0&\beta e^{i\phi_2^{(3)}}&\alpha e^{i\phi_3^{(3)}}
    \end{pmatrix},
    \label{eq:3band-U-Gamma1}
\end{equation}
where for convenience we made the $\Gamma_1$ dependence implicit. The real and positive coefficients $\alpha$ and $\beta$ are such that $\alpha^2+\beta^2=1$. Matrix $U$ also obeys one of the three following conditions,
\begin{equation}
    \alpha=0,\hspace{2mm}\beta=0,\hspace{2mm}e^{i(\phi_3^{(2)}-\phi_2^{(2)})}=-e^{i(\phi_3^{(3)}-\phi_2^{(3)})}.
    \label{eq:3band-3sol-Gamma1}
\end{equation}
These three solutions give rise to the three configurations of diagonal zeros displayed in Fig. \ref{fig:Zéros-3band-Gamma1}, respectively.

Notice that in the case of Fig. \ref{fig:Zéros-3band-Gamma1}c, the diagonal zeros $\zeta_{22}$ and $\zeta_{23}$ are located in the gap between $\epsilon_2$ and $\epsilon_3$. This situation only occurs in multiband systems, and is what gives rise to situations where Proposition \ref{Proposition-N-bandes} is not obeyed.

\begin{figure}[t]
    \centering
    \includegraphics[width=\linewidth]{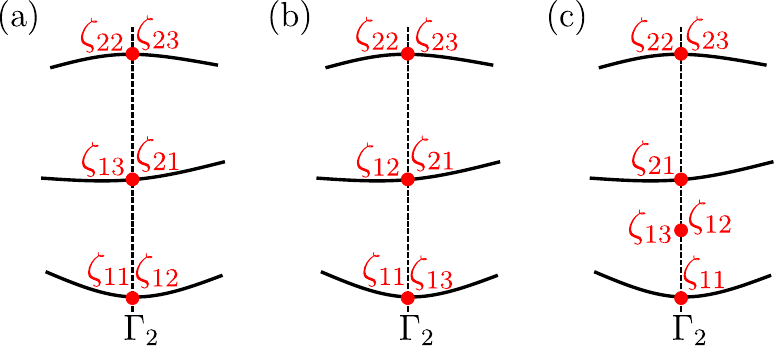}
    \caption{The three possible configurations of diagonal zeros at $\Gamma_2$. Figs. \ref{fig:Zéros-3band-Gamma2}a, \ref{fig:Zéros-3band-Gamma2}b and \ref{fig:Zéros-3band-Gamma2}c correspond to the first, second and third solutions in Eq.~(\ref{eq:3band-3sol-Gamma2}), respectively.}
    \label{fig:Zéros-3band-Gamma2}
\end{figure}

As for $\Gamma_2$, $U$ takes the following form,
\begin{equation}
    U=\begin{pmatrix}
        0&0&e^{i\phi_3^{(1)}}\\
        \alpha e^{i\phi_1^{(2)}}&\beta e^{i\phi_2^{(2)}}&0\\
        \beta e^{i\phi_1^{(3)}}&\alpha e^{i\phi_2^{(3)}}&0
    \end{pmatrix},
    \label{eq:3band-U-Gamma2}
\end{equation}
where the positive coefficients $\alpha$ and $\beta$ are also such that $\alpha^2+\beta^2=1$. We emphasize that $\alpha$ and $\beta$ correspond to $\Gamma_2$, and are different from the ones in Eq.~(\ref{eq:3band-U-Gamma1}), that correspond to $\Gamma_1$. Here, matrix $U$ in Eq.~(\ref{eq:3band-U-Gamma2}) must obey one of the three following conditions,
\begin{equation}
    \alpha=0,\hspace{2mm}\beta=0,\hspace{2mm}e^{i(\phi_1^{(3)}-\phi_1^{(2)})}=-e^{i(\phi_2^{(3)}-\phi_2^{(2)})}.
    \label{eq:3band-3sol-Gamma2}
\end{equation}
Similarly to $\Gamma_1$, a configuration of diagonal zeros can be obtained for each of the three conditions, as shown in Fig. \ref{fig:Zéros-3band-Gamma2}.
We thus have three possible configurations of diagonal zeros at each of the two TRIMs, and therefore nine possible behaviors. Let us focus on the combination of Fig. \ref{fig:Zéros-3band-Gamma1}a at $\Gamma_1$ and Fig. \ref{fig:Zéros-3band-Gamma2}c at $\Gamma_2$, represented in Fig. \ref{fig:Figure-Trois-Bandes}b. 

Between the third and second bands, we see that the diagonal zero $\zeta_{21}$ fully crosses the gap from $\Gamma_1$ to $\Gamma_2$, while $\zeta_{22}$ does so in the opposite way. As pictured in Fig. \ref{fig:Figure-Trois-Bandes}b, the two diagonal zeros must therefore cross at least once between the two TRIM points. Since the orbital $\ket{\phi_1}$ is of opposite parity as $\ket{\phi_2}$, the zero crossing is non-trivial and Proposition \ref{Proposition-N-bandes} is obeyed in the case $N_{\text{occ}}=2$. However, the situation changes when considering the case where $N_{\text{occ}}=1$. Indeed while $\zeta_{11}$ fully crosses the gap between $\Gamma_1$ and $\Gamma_2$, as pictured in Fig. \ref{fig:Figure-Trois-Bandes}b, the same does not happen for $\zeta_{12}$ and $\zeta_{13}$. The latter only cross the gap partially, and therefore the crossing between them and $\zeta_{11}$ does not fall in our definition of a zero-zero crossing as discussed in Sec. \ref{Sec:two-band-models}. In particular, the crossing can happen directly at $\Gamma_2$, near the energy of the band, which would make it impossible to observe. Consequently, there is no non-trivial zero-zero crossing between $\Gamma_1$ and $\Gamma_2$ for $N_{\text{occ}}=1$ despite the fact that the two TRIMs are of opposite parity. Proposition \ref{Proposition-N-bandes} is therefore not obeyed in this case. 

There are five such cases in three-band models where Proposition \ref{Proposition-N-bandes} does not hold \cite{SM}. All of them are in the configuration pictured in Fig. \ref{fig:Figure-Trois-Bandes}a. This configuration can be seen as the addition of a topologically trivial band to a topologically non-trivial two-band situation. Without the second band, Fig. \ref{fig:Figure-Trois-Bandes}a becomes a non-trivial two-band situation as described in Sec. \ref{Sec:two-band-models}, in the sense that $\Gamma_1$ and $\Gamma_2$ are of opposite parity and therefore have a non-trivial contribution to the $\mathbb{Z}_2$ topological invariants. In that sense the second band in Fig. \ref{fig:Figure-Trois-Bandes}a is trivial, since its parity eigenvalues will not contribute to to the relative parity of $\Gamma_1$ and $\Gamma_2$, and especially not to the $\mathbb{Z}_2$ topology. We therefore argue that the zeros surfaces thus display a fragile character, analogously to the band topology of fragile topological insulators \cite{Po2018}. We emphasize the limits of the analogy. First, the triviality of the second band in Fig. \ref{fig:Figure-Trois-Bandes}a is only with respect to $\Gamma_1$ and $\Gamma_2$, and is therefore partial. It is fully trivial only when all involved TRIMs are of the same band parity. Second, we use this analogy to encapsulate the idea that inserting a (partially) trivial band between two inverted bands breaks the ability of the zeros surface to determine the $\mathbb{Z}_2$ topology.

\section*{Conclusion}

We introduced the zeros surface, the set of $\bm{k}$-points where two zeros of different diagonal components of the Green's function in the orbital basis cross. In two-band systems, this zeros surface separates TRIM points with opposite parity in reciprocal space. It thereby allows one to directly, visually, calculate the four $\mathbb{Z}_2$ topological invariants. We illustrated the results with three topological phases of the Wilson-Dirac model.

In three-band models, the same holds in all possible cases but one: the case where one trivial band is added in between two inverted bands. There, the zeros of the Green's function do not necessarily cross the gap entirely. We associate this behaviour to fragile topological insulators, in which sense the zeros surface has a fragile character.

To prove this, we leveraged the zero-pole representation of the Green's function $G$, which states that the zeros of the $j$-th diagonal component of the Green's function, written in the orbital basis, are the eigenvalues of $[H]_{jj}$, the Bloch Hamiltonian without the $j$-th orbital $\ket{\phi_j}$. In this way, these diagonal zeros of $G$ are also interpreted as marginal band dispersions, in analogy with marginal probability distributions.

  Our study therefore establishes the Green function as a tool that offers a simple and visual way to determine the $\mathbb{Z}_2$ topological invariants in cases where two bands invert, without the computation of eigenvectors nor Wilson loops. Unlike in Chern insulators, there is to this date no direct way of experimentally measuring the $\mathbb{Z}_2$ topological invariants. Our findings could lead to an experimental determination of the $\mathbb{Z}_2$ topological invariants, 
  for example using ARPES measurements where light polarisation is leveraged to reconstruct the orbital character of the band structure \cite{Sahrakorpi2006,Yi2019,Louat2019}. In conjunction with the Kramers-Krönig relations, one could then in principle reconstruct the Green's function in the orbital basis \cite{Meevasana2008,Sobota2021,Shastry2024}. Although using ARPES to reconstruct the diagonal zeros is a theoretical possibility, its practical feasibility remains to be established \cite{Yam2025}.  One possible future direction of research is to see how the diagonal zeros relate to topology in interacting and non-crystalline sytems, as done in Ref. \cite{Tran2022}, where singular zeros also appear. 

\section*{Acknowledgments}
We thank Frédéric Piéchon, Mark O. Goerbig,  Adolfo G. Grushin and Jasper Van Wezel for stimulating discussions. We also thank Andrej Meszaros for constructive comments. F.S thanks financial support from the French National Research Agency (projects TWISTGRAPH and QMAHT) under Grants No. ANR-21-CE47-0018 and ANR-22-CE30-0032, respectively. 

\bibliography{Biblio2}

\end{document}